\documentclass[9pt,twocolumn,twoside]{opticajnl}
\journal{opticajournal} % use for journal or Optica Open submissions

\setboolean{shortarticle}{false} % True for OL & short Optica
\title{Fast spectroscopic imaging using extreme ultraviolet interferometry}
\author[1]{Hannah C. Strauch}
\author[2]{Fengling Zhang}
\author[1]{Stefan Mathias}
\author[3]{Thorsten Hohage}
\author[2]{Stefan Witte}
\author[1,*]{G. S. Matthijs Jansen}

\affil[1]{1\textsuperscript{st} Institute of Physics, University of Göttingen, Germany}
\affil[2]{Advanced Research Center for Nanolithography ARCNL, Amsterdam, the Netherlands}
\affil[3]{Institute for Numerical and Applied Mathematics, University of Göttingen, Germany}

\affil[*]{gsmjansen@uni-goettingen.de}

\begin{abstract}
Extreme ultraviolet pulses as generated by high harmonic generation (HHG) are a powerful tool for both time-resolved spectroscopy and coherent diffractive imaging. However, the integration of spectroscopy and microscopy to harness the unique broadband spectra provided by HHG is hardly explored due to the challenge to decouple spectroscopic and microscopic information. Here, we present an interferometric approach to this problem that combines Fourier transform spectroscopy (FTS) with Fourier transform holography (FTH). This is made possible by the generation of phase-locked pulses using a pair of HHG sources. Crucially, in our geometry the number of interferometric measurements required is at most equal to the number of high-harmonics in the illumination, and can be further reduced by incorporating prior knowledge about the structure of the FTH sample. Compared to conventional FTS, this approach achieves over an order of magnitude increase in acquisition speed for full spectro-microscopic data, and furthermore allows diffraction-limited computational imaging. 
\end{abstract}

\setboolean{displaycopyright}{false} % Do not include copyright or licensing information in submission.

\begin{document}

\maketitle

% \ifbool{shortarticle}{}{
\section{Introduction}
% }
Coherent extreme ultraviolet light produced by high-harmonic generation \cite{krausz_attosecond_2009} provides a unique opportunity for the study of ultrafast dynamics in condensed matter at the nanoscale. On the one hand, the short wavelength yields a favourable Abbe resolution limit, and extreme ultraviolet (EUV) coherent diffractive imaging nowadays enables imaging at nanometer-scale resolution \cite{loetgering_advances_2022, battistelli_coherent_2023, seaberg_ultrahigh_2011}. On the other hand, the wide range of photon energies and ultrashort, attosecond (as) to femtosecond (fs) pulse durations of EUV light pulses from high-harmonic generation (HHG) provide access to a wide range of elemental absorption edges. Here, time-resolved spectroscopy yields a sensitive element-resolved probe of dynamics of the electrons, the spins and the lattice in optically excited matter \cite{schultze_attosecond_2014, siegrist_light-wave_2019, geneaux_transient_2019, de_vos_ultrafast_2023, heinrich_electronic_2023, moller_verification_2024, probst_unraveling_2024}. The combination of both these approaches, namely nanometer-scale microscopic time-resolved EUV spectroscopy, is highly appealing, as it would enable the study of complex dynamics in nanoscale structures ranging from naturally-inhomogeneous quantum materials to fabricated heterostructures. However, such multi-dimensional measurements remain out of reach for HHG light sources as is, and can so far only be performed at significant experimental cost at accelerator-based EUV light sources \cite{johnson_quantitative_2021, johnson_ultrafast_2023}. 

Holography plays an important role in the development of advanced EUV microscopy \cite{mcnulty_high-resolution_1992, he_use_2004, marchesini_massively_2008, tenner_fourier_2014, geilhufe_achieving_2020, eschen_towards_2021}. For example, time-resolved EUV microscopy is commonly performed by combining Fourier-transform holography (FTH) with numerical phase retrieval \cite{zayko_ultrafast_2021, johnson_ultrafast_2023}. In FTH, the diffracted wave from the sample is interfered with a reference wave that is generated by a point-like structure (usually a pinhole in transmission geometry). The resulting fringe pattern in the far-field diffraction pattern allows for direct image reconstruction by a single Fourier transform. The single-shot nature of the FTH measurement is critical for the implementation of time-resolved studies, as it implies that only the pump-probe delay needs to be scanned during an experiment. In contrast, ptychography is a powerful imaging method that enables diffraction-limited resolution and can handle experimental challenges such as partial coherence, multi-spectral illumination and structured illumination profiles  \cite{thibault_high-resolution_2008, zhang_ptychographic_2016, loetgering_advances_2022}. To achieve this, ptychography relies on spatial scanning of the sample in overlapping steps such that the phase retrieval problem is sufficiently constrained. The need for overlap is significantly increased for multi-spectral illumination \cite{zhang_ptychographic_2016, loetgering_tailoring_2021, loetgering_advances_2022}. A full time- and spectrum-resolved ptychography measurement would therefore place significant requirements on the brightness and stability of the EUV light source. 
Thus, there is a clear need for a fast spectromicroscopy technique that reduces the data collection time for scanning in the spatial and spectral domains. 

\begin{figure}[th]
    \centering
    % [overview figure]
    \includegraphics[width=\linewidth]{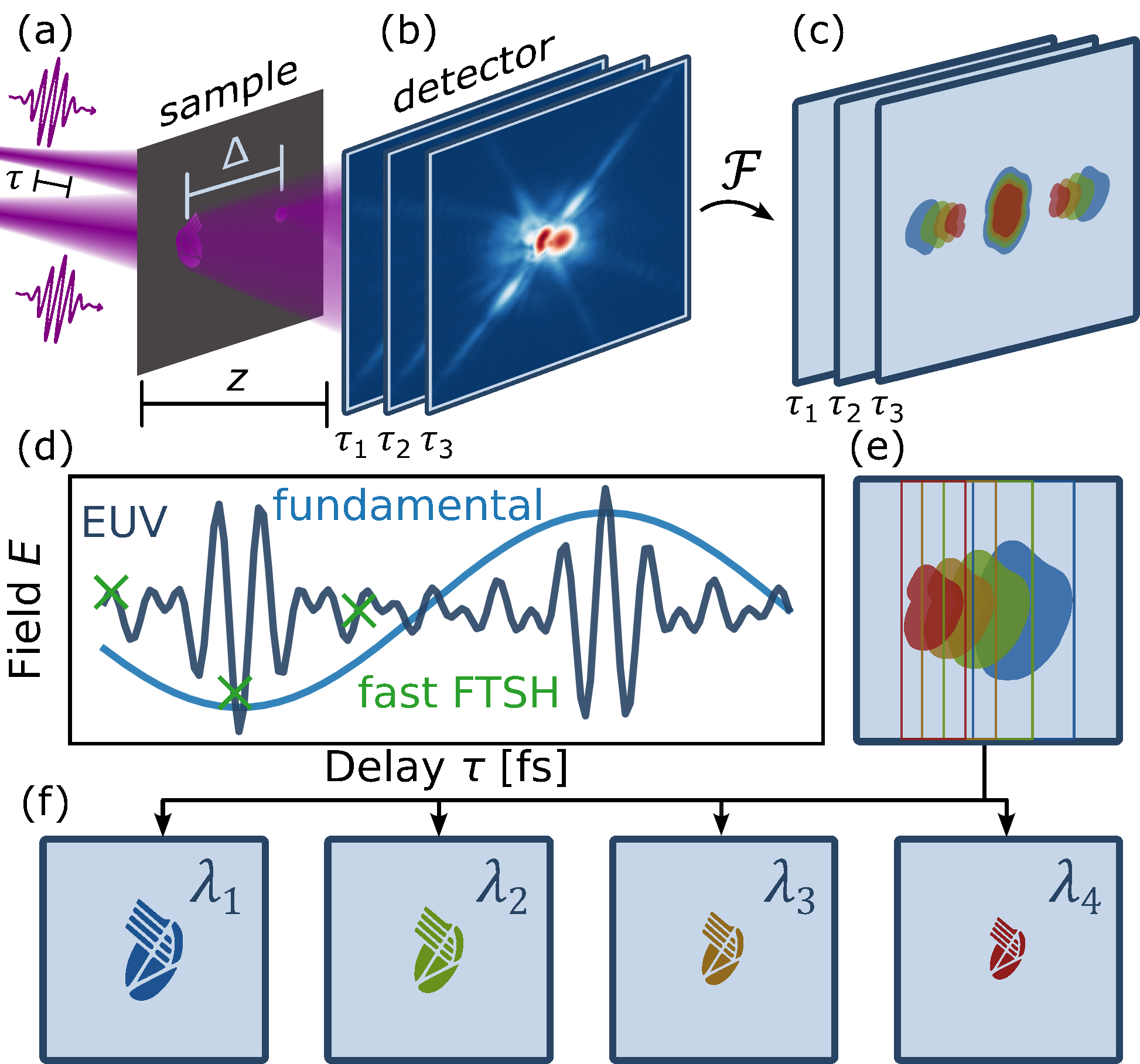}
    \caption{Overview of the Fourier transform spectroscopic holography (FTSH) workflow. (a) Two phase-locked polychromatic EUV beams from a pair of HHG sources illuminate the sample and reference structures. (b) The diffraction pattern is detected by a camera in the far-field. (c) As in conventional FTH, a single Fourier transform is used to analyze the interference pattern, however in this case the holograms of the individual photon energies are superimposed. (d) By time-shifting the reference wave, Fourier transform spectroscopy allows to measure the spectrum at each pixel. The necessary sampling for this measurement is significantly reduced when the multi-wavelength nature of the hologram (e) is considered, as the different spectral components are only partially overlapping. Thus, an HHG spectrum with $N_\lambda$ harmonics requires at most $N_\lambda$ and typically less measurements to fully recover a spectrally-resolved image (f).}
    \label{fig:overview_sfth}
\end{figure}

\begin{figure*}[ht!]
    \centering
    \includegraphics[width = 0.75\linewidth]{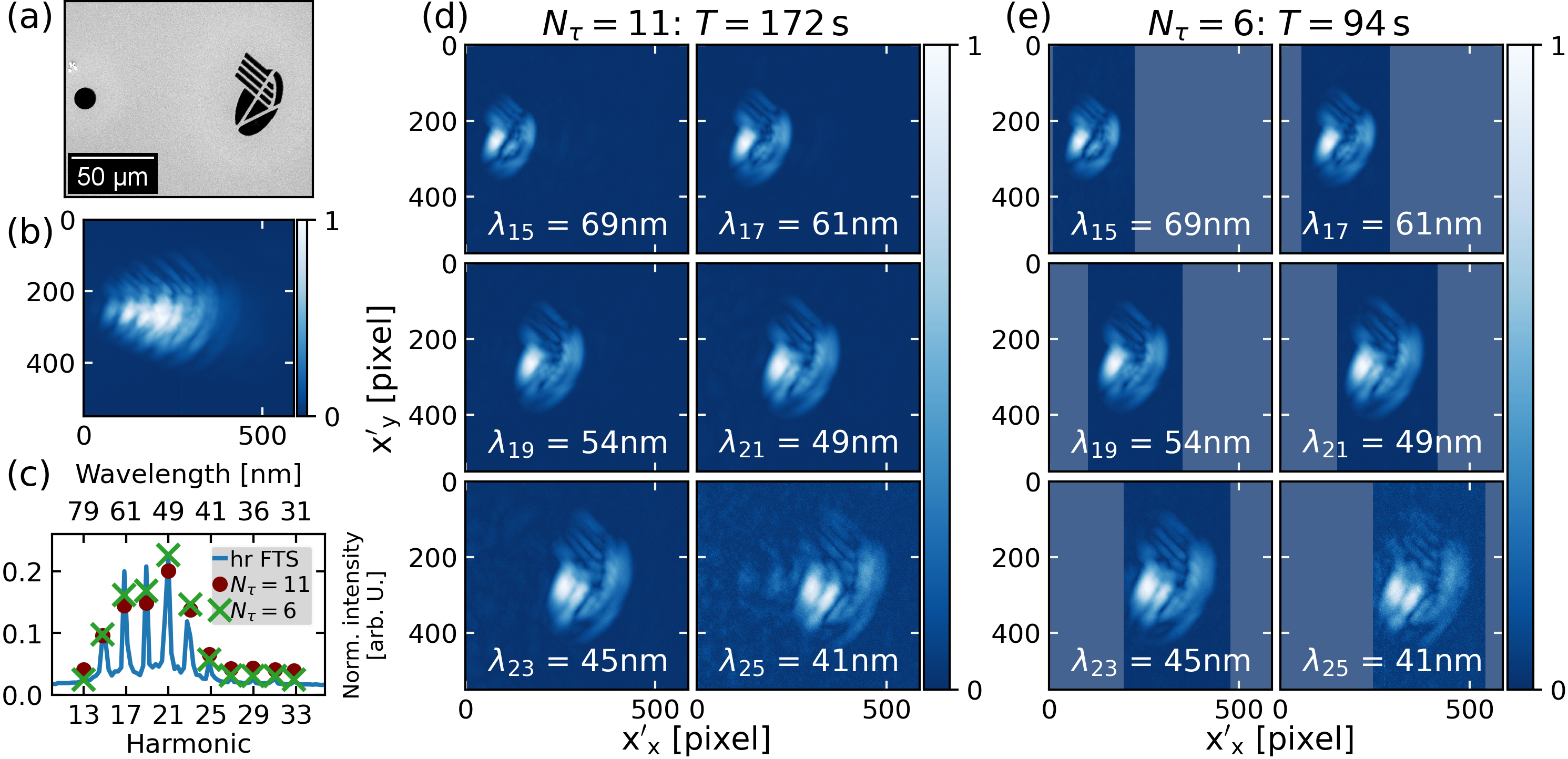}
    \caption{Fast Fourier-transform spectroscopic holography using minimal sampling. 
    a) Scanning electron microscopy image of the FTH sample, which was made by focused ion beam milling of a gold-coated silicon nitride membrane.
    b) Exemplary amplitude $|M(x',T)|$ of the multi-wavelength hologram, showing that the different spectral components are partially overlapping. 
    c) Extracted spectral intensity from the high resolution FTS scan, a dataset with $N_\tau = 11$ shots and $N_\tau = 6$ shots. The fundamental wavelength is 1030~nm. For comparison, the data was normalized to the total observed intensity. The small differences between the different datasets can be attributed to intensity drift and shot-to-shot variation in the HHG light source.
    d) Spectroscopic holography results from the $N_\tau = 11$ dataset for the 6 brightest high harmonics. 
    e) The corresponding spectroscopic holography results from the $N_\tau = 6$ dataset. 
    }
    \label{fig:few-shot_holograms}
\end{figure*}

In this Article, we demonstrate Fourier-transform spectroscopic holography (FTSH) (Fig.~\ref{fig:overview_sfth}), which combines FTH with an interferometric Fourier-transform spectroscopy measurement. For an HHG spectrum with $N_{\lambda}$ harmonics, FTSH requires at most $N_{\lambda}$ measurements and allows for further reduction by incorporating prior knowledge of the sample structure. In the experiment, we achieve flux-efficient FTSH by focusing the output of two phase-locked HHG sources \cite{jansen_spatially_2016} to a conventional FTH sample, such that the sample and reference structures are illuminated by separately controlled EUV pulses. In this geometry, we exploit the intrinsic coupling between spectral and spatial information in the diffraction pattern, and show that this enables a strong reduction of the FTS sampling requirements. For our HHG spectrum (covering up to 11 harmonics from 80 to 29~nm, see Fig.~\ref{fig:few-shot_holograms}b) and sample geometry, we find that only 6 phase steps are necessary to recover all spectral components completely.

% \ifbool{shortarticle}{}{
\section{Theory}
% }
Fourier-transform spectroscopic holography (Figure~\ref{fig:overview_sfth}) can be understood as an extension of FTH towards multiple wavelengths. For a single wavelength $\lambda_i$, the far-field diffraction pattern of an FTH sample can be expressed as 
% \begin{equation} 
$
    I_{\lambda_i}(k) = |\tilde{p}_i|^2 + |\tilde{r}_i|^2 + \tilde{p}_i \tilde{r}_i^* + \tilde{p}_i^* \tilde{r}_i, 
    % I_{\lambda_i}(k) = |\tilde{p}_i + \tilde{r}_i|^2, 
$
% \label{eq:FTHdiffraction}
% \end{equation}
where $\tilde{p}$ and $\tilde{r}$ are the electric fields at the detection plane due to the probed sample and reference, respectively, while $k$ is the spatial frequency of the scattered wave. The fields $\tilde{p}$ and $\tilde{r}$ can be related to the fields $p$ and $r$ at the sample plane by optical propagation. For Fraunhofer diffraction, the propagation is given by a scaled Fourier transform, where the scale (i.e., the extent of the diffraction pattern on the camera) is proportional to the wavelength of illumination. Consequently, a Fourier transform ($\mathcal{F}$)
\begin{equation}
% \begin{split}
    \mathcal{F}\{I_{\lambda_i}\}(x) =  \ p_i \ast p_i^* + r_i \ast r_i^* 
                                     + p_i \ast r_i^* + p_i^* \ast r_i
\label{eq:FTHbasic}
% \end{split}
\end{equation}
provides direct access to the holograms ($p_i \ast r_i^*$ and $p_i^* \ast r_i$) in the real-space coordinates $x$. It is well known that the resolution of these holograms is limited by the size of the reference $r$, and furthermore that the hologram appears separated from the autocorrelation terms ($p_i \ast p_i^*$ and $r_i \ast r_i^*$) at a distance $\Delta$ from the center that is equal to to the separation of $p$ and $r$. In terms of pixels of the reconstructed image, the separation becomes $\Delta_{\rm px} = 2 \Delta D / \lambda z$, where $z$ is the distance and $D$ the size of the detector. Notably, the numerical hologram separation $\Delta_{\rm px}$ is inversely proportional to the wavelength as a consequence of the wavelength scaling of the Fraunhofer propagation operator. 

While the precise position of the hologram in the frame is not important in monochromatic FTH, it becomes relevant for spectrally-resolved applications. If an FTH hologram is recorded using polychromatic radiation, a spatial Fourier transform of the data yields an image where the individual holograms are shifted according to the illuminating wavelength. This effect has been harnessed to perform single-shot spectrally resolved FTH using the individual high harmonics of a HHG light source \cite{williams_fourier_2015}. However, this requires a specific experimental geometry: in order to fully separate the individual holograms, the sample width needs to be reduced significantly (or equivalently the $\Delta$ must be increased) by a factor that depends on the wavelength difference of neighbouring high harmonics. This drastically reduces the achievable field of view and limits the spectral resolution, altogether preventing the application to continuous (attosecond) EUV spectra. Therefore, another method is necessary to resolve the individual spectral components. 

Fourier transform spectroscopy (FTS) is a powerful method that provides such a capability: analogous to FTH, interference of two pulses delayed in the time domain gives access to the spectrum by an inverse Fourier transform. FTS-based methods are already commonly used to perform hyperspectral imaging in the visible and infrared ranges (e.g., \cite{kalenkov_hyperspectral_2017, genco_k-space_2022}). In the last decade, FTS at extreme ultraviolet wavelengths has been enabled by the generation of phase-locked EUV pulse pairs by phase-locked laser pulses \cite{jansen_spatially_2016, meng_octave-spanning_2015}. Integrated in a coherent diffractive imaging experiment, spatially-resolved FTS yields monochromatic diffraction patterns that can be used for numerical image reconstruction \cite{jansen_diffractive_2018}. In order to satisfy the Nyquist-Shannon sampling theorem and to accurately measure the complete spectrum, however, typically a few hundred interferometric diffraction patterns at different delays must be measured. 
% However, the extension of this method to time-resolved studies is limited by the need for long pulse delay scans to resolve the full spectral information. 
%and so far low photon efficiency due to the need for large, mm-scale EUV beams.

Here, we show that the individual limitations of FTH and FTS can be lifted when they are combined in a HHG-based interferometric measurement (cf. Fig.~\ref{fig:overview_sfth}). In particular, we propose to combine the FTSH measurement with prior knowledge of the experimental parameters. The generated wavelengths of a HHG source are generally known in advance to a good accuracy, e.g., from a prior spectroscopy measurement or from a measurement of the fundamental laser parameters. This knowledge can already significantly reduce the sampling requirements, but a further reduction is possible: Combined with knowledge of the FTH mask structure, it is possible to calculate which spectral components can contribute to each pixel in the FTH hologram. For comb-like HHG spectra spanning an octave or more, it is generally found that only a small number of harmonics contribute to each pixel in the multi-wavelength hologram.
% this leads in most cases to the situation that at each position several monochromatic holograms overlap, but the number of overlapping components is typically much smaller than the number of high harmonics $N_{\lambda}$.% in the illumination. 

It is instructive to consider the following example: an FTH sample containing a reference structure with negligible width and a sample with width $W$ (along the sample-reference direction) with a sample-reference separation $\Delta = 2 W$ is illuminated by a HHG spectrum containing the odd harmonics 13 to 29 of an 1030 nm driving laser. In this case, the hologram of the 29\textsuperscript{th} harmonic has significant overlap with the 27\textsuperscript{th}, successively less overlap with the 25\textsuperscript{th}, 23\textsuperscript{rd} and 21\textsuperscript{st} harmonic holograms, and no overlap with holograms of the 19\textsuperscript{th} and lower high harmonics. In fact, no more than 5 high harmonics overlap at any point in the multi-wavelength hologram. This suggests that the sampling requirements in an FTS measurement can be reduced. 

In FTSH, the reference beam is delayed by a time $\tau$ with respect to the probe beam. This leads to a measured intensity given by $ I(k,\tau) = \sum_i^{N_{\lambda}}|\tilde{p}_i + \tilde{r}_i e^{-i 2\pi c \tau/ \lambda_i}|$. By applying a Fourier transform, we can express the measured multi-wavelength hologram $M(x',T)$ as
\begin{equation}
\begin{split}
    M(x',\tau) %= \sum_i^N_\lambda \mathcal{F}(I_{\lambda_i})(r)
    = \sum_i^{N_{\lambda}} & \ p_i \ast p_i^* + r_i \ast r_i^* 
                 + (p_i \ast r_i^*) e^{-i 2\pi c \tau/ \lambda_i} + c.c.,
                  % + (p_i^* \ast r_i) e^{-i 2\pi c \tau/ \lambda_i}, 
% M(x',T) & = \sum_i^{N_\lambda} \mathcal{F}(I_{\lambda_i})(x') \\
%        & = \ldots{} + \sum_i^{N_\lambda} (p_i^* \ast r_i) e^{-i 2\pi c \tau/ \lambda_i} 
\end{split}
\label{eq:FTSH}
\end{equation}
where $c.c.$ indicates the complex conjugate and $c$ is the speed of light. 
% where we have written out only one of the cross-correlation terms and left out the constant autocorrelation terms for simplicity. $N_\lambda$ indicates the number of wavelengths in the illumination and $c$ is the speed of light. 
The coordinate $x'$ is Fourier conjugate to the pixel coordinates of the camera, and can be related to the real-space coordinate by considering the wavelength-dependent resolution and shift $\Delta_{\rm px}$. It deserves emphasis that the \emph{complex-valued} data $\sum_i^{N_\lambda} (p_i \ast r_i^*) e^{-i 2\pi c \tau/ \lambda_i}$ can be directly retrieved from the diffraction pattern measured at a single delay $\tau$ by spatially isolating the multi-wavelength hologram (see Fig.~\ref{fig:overview_sfth}c, e). 

The spectroscopic reconstruction problem can therefore be posed as follows: Given $M(x',\tau_j)$ for $j \in [1,N_\tau]$, find the complex-valued $p_i^* \ast r_i$ for each $i\in [1,N_\lambda]$ at each $x'$. For suitably chosen delays $\tau_j$, this problem is well posed, and can be solved by matrix inversion or least-squares methods. Ignoring prior knowledge of the shape of the holography sample there are at most $N_\lambda+1$ unknown complex values for each $x'$, indicating that at most $N_\tau = N_\lambda +1$ measurements are required. This is reduced by 1 for sufficiently large $\Delta$, since the autocorrelation terms can be filtered out spatially. A more dramatic reduction is achieved if the partial overlap of the monochromatic holograms is considered, as explained by the previous example. 

% \begin{figure}[htb!]
%     \centering
%     \includegraphics[width = \linewidth]{images/overview6shown1colV6.png}
%     \caption{Fast Fourier-transform spectroscopic holography using minimal sampling. 
%     a) Scanning electron microscopy image of the FTH sample, which was made by focused ion beam milling of a gold-coated silicon nitride membrane.
%     b) Exemplary amplitude $|M(x',T)|$ of the multi-wavelength hologram, showing that the different spectral components are partially overlapping. 
%     c) Extracted spectral intensity from the high resolution FTS scan, a dataset with $N_\tau = 11$ shots and $N_\tau = 6$ shots. The fundamental wavelength is 1030~nm. The small differences between the different datasets can be attributed to intensity drift and shot-to-shot variation in the HHG light source.
%     d) Spectroscopic imaging results from the $N_\tau = 11$ dataset for the 6 brightest high harmonics. 
%     e) The corresponding spectroscopic imaging results from the $N_\tau = 6$ dataset. 
%     }
%     \label{fig:few-shot_holograms}
% \end{figure}

% \ifbool{shortarticle}{}{
\section{Spectroscopic holography}
% }
In order to record FTSH data at EUV wavelengths, we employ an ultrastable birefringent common-path interferometer \cite{brida_phase-locked_2012, jansen_spatially_2016} to split the output of a 1~kHz, 35~fs, 1030~nm laser (Light Conversion Pharos, compressed by cascaded nonlinear compression in 800~mbar argon \cite{tsai_nonlinear_2022}) into phase-locked pulses with a controllable delay. By tilting one wedge of the common-path interferometer, these pulses are focused to two separate spots in a Krypton gas cell, where they generate two phase-locked EUV beams. After a 200~nm Al filter to block the fundamental laser light, the high-harmonic spectrum of both pulses spans from 80 to 29~nm. At 94~cm after the HHG, we use a broadband curved multilayer mirror ($f = 25$~cm) to image the pulse pair onto the sample region. In the focus plane, the two EUV beams are separated by roughly 70~\textmu{}m, and the diameter of the individual beams is roughly 25~\textmu{}m. To facilitate a more homogeneous illumination of the sample, we place the FTH sample $\approx$5~mm downstream of the focus. The EUV camera (Andor IkonL, 13.5~micron pixel size) is placed 10~cm behind the sample.    

In this geometry, we now record full FTSH data. Analysis of a high-resolution FTS scan with a time-step of 23~as over a range of 11.6~fs %, i.e., 3.3 periods of the fundamental, 
confirms that the HHG spectrum consists of 11 high harmonics with frequencies that match to the odd harmonics 13 to 33. The oscillation period of these harmonics ranges from 262~as to 103~as. In the following, this dataset will serve as a reference and allows to benchmark the analysis results when only using few-delay subsets of the full data. 

% Next, we subsample the full FTS scan in order to demonstrate the sampling requirements for full spectral resolution. By the Nyquist theorem, the spectral resolution of a Fourier transform of data from $N_\tau = N_j = 11$ equidistant delays (stepsize 190~as) covering a half cycle of the driving laser pulse, exactly matches with the spacing of the high harmonics, namely twice the fundamental frequency. However, we emphasize that the spectral amplitudes cannot be reconstructed using a Fourier transform at this minimal sampling. Instead, we solve \eqref{eq:FTSH} using the Newton method implemented in RegPy \textcolor{red}{cite{}}, a Python-based toolbox for implementing and solving (potentially) ill-posed problems. While the spectral reconstruction can also be achieved using other methods, this method provides a number of advanced capabilities that we will exploit later. 
Next, we therefore subsample the full FTS scan in order to demonstrate the minimum sampling requirements for full spectroscopic image reconstruction. Specifically, we now consider interferometric diffraction patterns at $N_\tau = N_\lambda = 11$ equidistant delays in the range 0 - 1.7~fs (step size 190~as), which corresponds to a half cycle of the driving laser pulse. By the Nyquist theorem, the spectral resolution of the Fourier transform of this data exactly matches with the spacing of the high harmonics, namely twice the fundamental frequency. However, the sampling frequency is much lower than the highest frequency in the data, and we emphasize that the spectral amplitudes cannot be reconstructed using a Fourier transform. Instead, we solve \eqref{eq:FTSH} using the Newton conjugate-gradient method implemented in RegPy \cite{hanke_regularizing_1997, Maretzke_regularized_16}, a Python-based toolbox for implementing and solving (potentially ill-posed) inverse problems. While the spectral reconstruction can also be achieved using other methods, this toolbox provides a number of advanced capabilities that we will exploit later. % The data and code for this analysis are publicly available in Ref~\textcolor{blue}{[citation follows]}.

For the case that $N_\tau = N_\lambda$, the reconstruction problem is fully constrained: for each pixel $x'$ we have $N$ complex-valued measurements and use these to determine $N$ complex-valued amplitudes. This is also reflected in the measurement results: as shown in Fig.~\ref{fig:few-shot_holograms}b-c, the $N_\tau = 11$ measurement allows to accurately extract the spectrally-resolved holograms from the data. This method already provides a powerful advantages over the full FTS scan: As a full FTS scan of the diffraction pattern requires to sample the highest harmonic with 2 points per oscillation and must also resolve the individual harmonics, the number of samples must be at least twice the order of the highest harmonic (i.e., 66 measurements for our spectrum with harmonics up to the 33\textsuperscript{rd} order), the total number of measurements is dramatically reduced. In practice, the sampling advantage is around an order of magnitude, as full FTS measurements in the EUV typically use hundreds of delays. In addition to a shorter measurement time, this implies that the short scan is much less sensitive to experimental drift, such as in the EUV beam pointing. 

% Still, it is possible to achieve a full reconstruction with even less measured delays. Observing the weak intensity of the highest and lowest harmonics, one approach is to exclude these harmonics from the analysis, and we thus perform a spectral analysis of $N_{\lambda}=9$ harmonics using $N_\tau=9$ delays. For the given HHG spectrum, this introduces an error below \textcolor{red}{XX\%}. 
Next, we exploit the intrinsic wavelength-sensitivity of FTH, which leads to the relative displacement of the individual spectral components that was discussed earlier. Based on the known width (30~\textmu{}m) and separation (100~\textmu{}m) of the sample and reference, we construct a simple mask to constrain which spectral components can contribute to each pixel $x'$. From this analysis, we find that at no point more than 6 high harmonics contribute to the signal (and 99\% of the pixels include contributions of $\leq5$ harmonics only). We therefore set $N_\tau = 6$ and use interference patterns spaced in delay by $1.7/5 = 0.34$~fs to again cover half cycle of the driving laser pulse. The mask can be implemented conveniently into the RegPy reconstruction method. As shown in Fig.~\ref{fig:few-shot_holograms}d, this approach enables spectroscopic imaging at a total measurement time of only 94\,s. % 1m 34s. % Note how the weak harmonics get assigned less intensity, as visible from the comparison of the spectra, because leakage from adjecent harmonics is suppressed by the mask.

\begin{figure}[hbt]
    \centering
    \includegraphics[width = \linewidth]{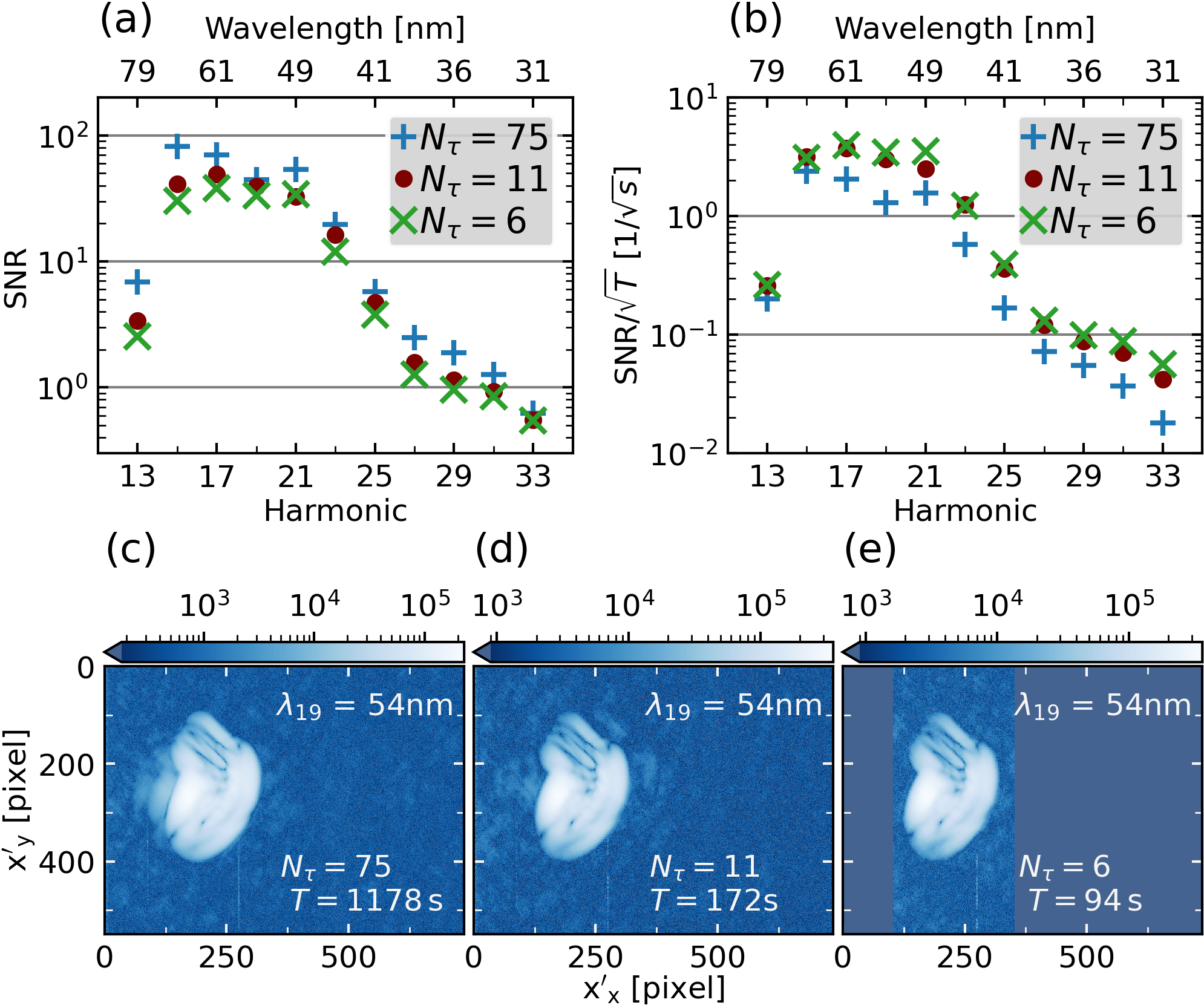}
    \caption{Comparison of image quality for minimally sampled FTSH compared to FTS with $N_\tau=75$ covering a delay range of 3.48~fs. This is the minimum delay range that allows to separate the individual high harmonics by a Fourier transform. (a) Experimental signal-to-noise ratio (SNR), calculated by comparing the average logo intensity to the background. (b) SNR scaled by $1/\sqrt{T}$, where $T$ is the total measurement time. (c-e) Exemplary spectrally-resolved hologram for the 19\textsuperscript{th} harmonic at 54~nm. 
    % (d) Visualization of the selected delay steps, overlaid on the fundamental $\lambda = 1030$~nm oscillation. 
    Although the overall best SNR is achieved from the long integration, the SNR is only $\approx$2x better than that of the $N_\tau=11$ analysis, while the total exposure time was 7x longer. Assuming that the SNR of the minimally sampled reconstructions is limited by photon shot noise, minimally sampled FTSH is expected to yield a 2x higher SNR than the full FTS measurement for equal data acquisition time. }
    \label{fig:snr_analysis}
\end{figure}

% \section{Discussion}
These results indicate that the sampling requirements in spatially-resolved Fourier transform spectroscopy, specifically when applied in combination with Fourier transform holography, is not determined by the well-known Nyquist-Shannon theorem applied to the HHG source spectrum, but it is rather determined by the total number of unknowns, which is in turn dependent on the FTH sample structure, the spectral resolution, and the bandwidth of the EUV spectrum. For typical HHG spectra and FTH sample structures, this leads to a significant reduction of the sampling requirements. In practice, the number of delay steps needed can be controlled by the separation~$\Delta$. Increasing $\Delta$ at fixed sample width leads to a reduced overlap of the individual spectral components and consequentially a reduced number of steps. However, the increased $\Delta$ also implies that a larger total field of view must be considered in FTH. For diffraction-limited imaging at high numerical aperture (see below), this can also result in a reduction of the final image resolution. Thus, in general a trade-off between image acquisition speed, image quality and experimental parameters such as the illumination beamline must be found. 

To quantify the benefit of minimal sampling FTSH compared to high-resolution FTS, we carry out an analysis of the effective signal-to-noise (SNR) ratio of the reconstructed images in Fig.~\ref{fig:snr_analysis}. Overall, we find that a longer total integration time leads to an increase in the SNR, which agrees with expectation, because the effects of shot-noise and shot-to-shot HHG variations are reduced. However, the improvement from $N_\tau=11$ to the full FTS result is disproportional to the increase in measurement time. Assuming that the noise is limited by shot-noise, the image quality can be compared if the SNR is scaled by $1/\sqrt{T}$, where $T$ is the total measurement time (see Fig.~\ref{fig:snr_analysis}c). From this analysis, we find that both minimally-sampled FTSH methods (11 delays without mask and 6 delays with applied mask, respectively) perform on average 2x better than the full FTS measurement. We tentatively attribute this difference to a reduced sensitivity to drifts in the EUV spectrum and beam pointing for the shorter FTSH measurement. 

The presented FTSH method relies on prior knowledge of the high-harmonic frequencies. Although these can be commonly estimated from the spectrum of the near-infrared driving pulse, the high-harmonic spectrum can be shifted due to various effects such as pulse chirp and intensity blue-shift \cite{chang_temporal_1998, shin_nonadiabatic_2001}. Therefore, we have performed an analysis of frequency miss-estimation on the reconstructed electric fields, from which we find that a $\pm 1$\% error in the fundamental frequency can lead to a root-mean-square difference up to 20\%. Such errors can be completely avoided, however, by performing an extended measurement with at least one extra delay. The number of independent pixels is normally much larger than the number of spectral components ($10^5$ pixels and $11$ frequency components in our work), and thus this extra image provides enough data to determine the high-harmonic frequencies precisely through iterative minimization. With regard to the measurement calibration, also sub-wavelength accuracy is required in the determination of the delays. This can for example be achieved using a stable interferometer \cite{brida_phase-locked_2012} or through separate interferometric delay calibration \cite{meng_interferometric_2016}. An advantage of the minimally-sampled FTSH scheme for typical HHG spectra is that the necessary delay range covers less than one micrometer, thereby reducing the requirements on the interferometer. 
% Especially with respect to the frequencies, this can be related by the choice of a short delay window of 1.7~fs: in this time window a $\pi/4$ phase shift requires a frequency shift of $\lambda_0/2$. 

\begin{figure}[htb]
    \centering
    \includegraphics[width=\linewidth]{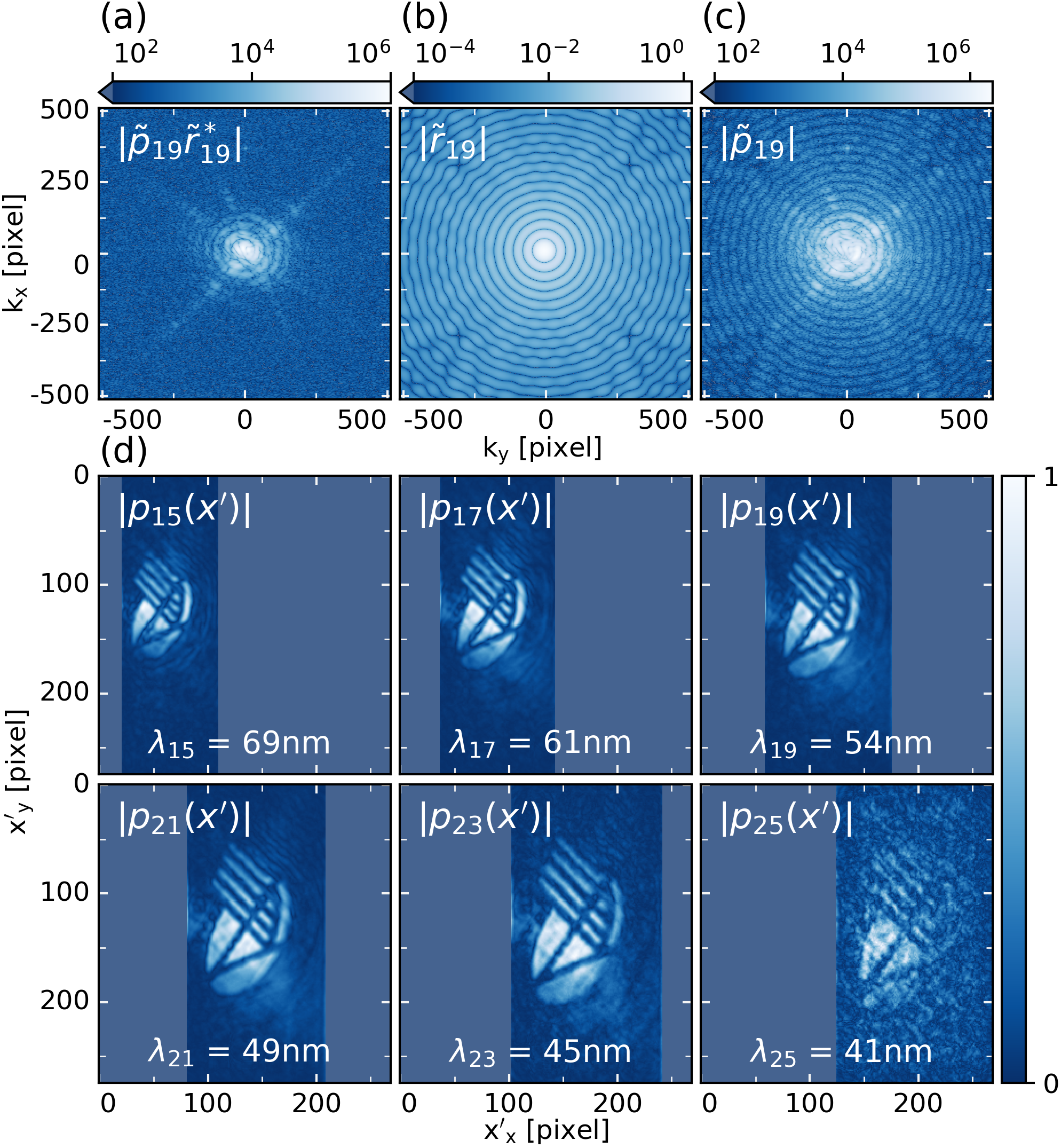}
    \caption{Diffraction-limited spectroscopic imaging based upon the $N_\tau = 6$ dataset with a total measurement time of 1m 34s. (a) Far-field interference pattern $|\tilde{p}_\lambda\tilde{r}_\lambda^*|$ of the 19\textsuperscript{th} harmonic at 54~nm. (b) The far-field diffraction pattern (Airy pattern) of the reference structure at 54~nm. (c) Reconstructed far-field diffraction pattern $|\tilde{p}_\lambda|$ of the object, as extracted from (a). (d) Reconstructed images for the 15\textsuperscript{th} to 25\textsuperscript{th} harmonics, respectively.}
    \label{fig:deconv}
\end{figure}

% \ifbool{shortarticle}{}{
\section{Diffraction-limited spectromicroscopy}
% }
A limitation of FTH is the trade-off between resolution (requiring small reference structures) and signal strength (requiring large references). Hence, many FTH experiments use subsequent iterative phase retrieval to improve the spatial resolution and SNR \cite{capotondi_scheme_2012, tenner_fourier_2014, geilhufe_achieving_2020}. In the following, we will demonstrate how iterative image reconstruction methods can be used to achieve diffraction-limited resolution in FTSH. In a first step, it is useful to separate the delay-dependent and delay-independent contributions to the diffraction pattern.
The delay-independent part has two contributions: $\sum_i^{N_\lambda} |\tilde{p}_i|^2$ and $ \sum_i^{N_\lambda} |\tilde{r}_i|^2$. These terms each represent a conventional broadband diffraction pattern, for which it is known that image reconstruction is possible with strong prior knowledge of the spectral response of the object \cite{huijts_broadband_2020}. Reconstruction of the individual spectral components, however, is not generally possible. Also, the incoherent sum of both broadband diffraction patterns will complicate image reconstruction. Consequentially, we will focus in the following on the spectrally-resolved interference patterns that are retrieved from the FTSH analysis.

After applying a Fourier transform to the output of the FTSH analysis (such as shown in Fig.~\ref{fig:few-shot_holograms}d, e), the monochromatic interference-diffraction patterns can be expressed as $m_\lambda(k) = \tilde{p}_\lambda\tilde{r}_\lambda^*(k)$. As this data already includes the phase, it is not necessary to implement phase retrieval algorithms as in monochromatic coherent diffractive imaging. %This data differs from the data that is used for monochromatic coherent diffractive imaging, which are given by $|\tilde{p}_\lambda|^2$. It is thus clear that the phase retrieval algorithms used for CDI cannot be applied. 
Instead, the image reconstruction of $p_\lambda$ from $m_\lambda$ can be posed as a deconvolution problem, where the point-spread function of the reference ($r_\lambda$) needs to be subtracted from the data. This can also be observed in the extracted monochromatic far-field interference pattern (Fig.~\ref{fig:deconv}a), where a clear imprint of the Airy pattern due to the 12~micron reference can be seen. We emphasize that the clear visibility of the Airy rings is due to the direct illumination of both the reference and the sample using focused EUV beams. The shape of the observed Airy pattern matches well to the expected pattern (Fig.~\ref{fig:deconv}b).

To reconstruct the diffraction-limited image from the monochromatic interferogram $m_\lambda(k)$ and the known reference $r_\lambda$, we use the Newton conjugate-gradient method to iteratively minimize $||\tilde{p}_\lambda\tilde{r}_\lambda^*(k) - m_\lambda(k)||$. In order to prevent overfitting to the noise, we use Morozov's discrepancy principle, i.e. we terminate the algorithm as soon as the residuals are on the order of the noise floor. This is commonly achieved after 11 Newton steps.
% In order to prevent overfitting to the noise, we terminate the algorithm as soon as the residuals are on the order of the noise floor, which is commonly achieved within 11 Newton steps. 
Using this procedure, we find accurate image reconstructions for up to 6 high-harmonics, see Fig.~\ref{fig:deconv}. Compared to the initial holograms that were limited by the 12~micron diameter reference, this enables a dramatic increase in the resolution to a final resolution of approximately 1.5~micron. This resolution corresponds to the numerical aperture of the observed reference diffraction. In following experiments, the experimental resolution can further be improved by the implementation of smaller or structured references, which can be illuminated efficiently because the EUV beams are focused to the reference and sample individually. Also, the resolution and SNR can be improved by averaging of several consecutive measurements (see supplementary information). It is also worth noting that this image reconstruction procedure does not depend on knowledge of the object support.

In summary, we have investigated the potential of Fourier-transform spectroscopic holography (FTSH), an interferometric technique that allows minute-scale spectromicroscopy at extreme ultraviolet photon energies. In the FTH geometry, prior knowledge of the illumination wavelengths allows a dramatic reduction in the sampling requirements. We have identified two sampling strategies for minimally sampled FTSH: (i) without assuming prior knowledge on the overlap of monochromatic holograms, the number of measurements can be reduced to the number of wavelength components in the illumination, and (ii) by incorporating such prior knowledge, a further reduction can be achieved. For our HHG spectrum and FTH sample, this reduces the number of measurements to 6, which is a typical value. Overall, this approach enables an order of magnitude reduction in the required sampling compared to full FTS sampling by the Nyquist-Shannon theorem, thereby speeding up the measurement dramatically. In the spectrally resolved holograms, we find that a higher SNR is achieved in equivalent time. This is made possible by the shorter measurement duration that reduces the effect of long-term drift and instability in the experimental setup. Finally, although the measurement is based on holography and intrinsically measures the convolution between sample and reference, FTSH does enable diffraction-limited resolution. We expect that this method will contribute strongly to the implementation of table-top extreme ultraviolet spectromicroscopy, for example in time-resolved experiments where a fast measurement is necessary to facilitate systematic studies.

\begin{backmatter}
\bmsection{Funding} This work was funded by the Deutsche Forschungsgemeinschaft (DFG, German Research Foundation) - 432680300/SFB 1456, projects B01 and C03. F.Z. and S.W. acknowledge support from the European Research Council (ERC-CoG project 3D-VIEW, 864016). 

The experiments were performed at ARCNL, a public-private partnership between the University of Amsterdam (UvA), Vrije Universiteit Amsterdam (VU), Rijksuniversiteit Groningen (RUG), the Dutch Research Council (NWO), and the semiconductor equipment manufacturer ASML.

% \bmsection{Acknowledgments} 
% The section title should not follow the numbering scheme of the body of the paper. Additional information crediting individuals who contributed to the work being reported, clarifying who received funding from a particular source, or other information that does not fit the criteria for the funding block may also be included; for example, ``K. Flockhart thanks the National Science Foundation for help identifying collaborators for this work.''

\bmsection{Disclosures} The authors declare no conflicts of interest.

% \bmsection{Data availability} Data underlying the results presented in this paper are not publicly available at this time but may be obtained from the authors upon reasonable request.

%\bmsection{Data availability} 
%The data and python code will be made publicly available upon publication of this article. 

% Data underlying the results presented in this paper are available in Ref. [3].
% The data underlying our simulation results can be generated using the software package that is available here [ref]

% \bmsection{Supplemental document}
% See Supplement 1 for supporting content. 

\end{backmatter}

% Bibliography
\bibliography{2024_EUV_Holo}

% Full bibliography added automatically for Optics Letters submissions; the following line will simply be ignored if submitting to other journals.
% Note that this extra page will not count against page length
\bibliographyfullrefs{2024_EUV_Holo}

\end{document}